\documentclass[aps,prl,twocolumn,superscriptaddress]{revtex4-2}
\usepackage{graphicx}%
\usepackage{multirow}%
\usepackage{amsmath,amssymb,amsfonts}%
\usepackage{amsthm}%
\usepackage{mathrsfs}%
\usepackage[title]{appendix}%
\usepackage{xcolor}%
\usepackage{textcomp}%
\usepackage{booktabs}%
\usepackage{algorithm}%
\usepackage{algorithmicx}%
\usepackage{algpseudocode}%
\usepackage{listings}%
\usepackage{etoolbox}
\usepackage[english]{babel}

\begin{document}

\title{Formation of C-centers in Si-based systems by light ion irradiation}


\author{C. Crosta}
\email[]{c.crosta3@campus.unimib.it}
\affiliation{Dipartimento di Scienza dei Materiali, Universit\`a di Milano-Bicocca and BiQute, via R. Cozzi, 20125 Milano, Italy}

\author{R. Nardin}
\affiliation{Dipartimento di Scienza dei Materiali, Universit\`a di Milano-Bicocca and BiQute, via R. Cozzi, 20125 Milano, Italy}

\author{P. Daoust}
\affiliation{Department of Engineering Physics, École Polytechnique de Montr\'eal, Montr\'eal, C.P. 6079, Succ. Centre-Ville, Montr\'eal, Qu\'ebec, Canada H3C 3A7}

\author{S. Achilli}
\affiliation{Dipartimento di Scienza dei Materiali, Universit\`a di Milano-Bicocca and BiQute, via R. Cozzi, 20125 Milano, Italy}

\author{I. Colombo}
\affiliation{Dipartimento di Scienza dei Materiali, Universit\`a di Milano-Bicocca and BiQute, via R. Cozzi, 20125 Milano, Italy}

\author{M. Campostrini}
\affiliation{Laboratori Nazionali di Legnaro, Istituto Nazionale di Fisica Nucleare (INFN-LNL), Viale dell’Universit\`a 2,
35020 Legnaro, Padua, Italy}

\author{E. Bonera}
\affiliation{Dipartimento di Scienza dei Materiali, Universit\`a di Milano-Bicocca and BiQute, via R. Cozzi, 20125 Milano, Italy}

\author{J. Pedrini}
\affiliation{Dipartimento di Scienza dei Materiali, Universit\`a di Milano-Bicocca and BiQute, via R. Cozzi, 20125 Milano, Italy}

\author{O. Moutanabbir}
\affiliation{Department of Engineering Physics, École Polytechnique de Montr\'eal, Montr\'eal, C.P. 6079, Succ. Centre-Ville, Montr\'eal, Qu\'ebec, Canada H3C 3A7}

\author{V. Rigato}
\affiliation{Laboratori Nazionali di Legnaro, Istituto Nazionale di Fisica Nucleare (INFN-LNL), Viale dell’Universit\`a 2,
35020 Legnaro, Padua, Italy}

\author{F. Pezzoli}
\email[]{fabio.pezzoli@unimib.it}
\affiliation{Dipartimento di Scienza dei Materiali, Università di Milano-Bicocca BiQute and INFN-LNL, Via R. Cozzi, 55, Milano, 20125, Italy}


\begin{abstract}
Atomic-scale crystal defects in Si are quantum-light sources offering tantalizing integration with existing photonic technologies. Yet, the controlled creation of near-infrared color centers for long-haul quantum communication and information still remains a challenge. In this work, we utilize light ions, such as $\mathrm{H^+}$ and $\mathrm{He^+}$, to gently generate quantum emitters in a crystalline Si matrix. Temperature-dependent photoluminescence measurements demonstrate the presence of optically-active defects, whose fluorescence matches the primary telecom window around 1550 nm. In addition, time-resolved investigations unveil long-lived excitonic states in the $\mu$s regime, thus confirming the formation of interstitial oxygen-carbon complexes, termed C-centers. Finally, we explored controlled ion irradiation strategies to seamlessly generate C-centers also in Ge-on-Si heterostructures, which offer an advanced technological platform for the future realization of integrated quantum photonics.
This analysis, informed by practical color center synthesis and proof-of-principle experiments in epitaxial architectures, indicates intriguing prospects and profitable strategies to advance the burgeoning field of light-based quantum technologies.
\end{abstract}


\maketitle

\section{Introduction}\label{sec1}

Quantum technologies look poised to revolutionize computing, communications, and sensing. Different approaches involving superconductors, trapped ions, neutral atoms, or spins have been reported as prospective platforms for quantum applications. While an important advantage of light-based quantum information processing is scalability, one of its main technological challenges is the implementation of single photon emitters operating in the telecom range \cite{andrini2024solid, aberl2024all}. In this context, the so-called L-band (1565 - 1625 nm \cite{cannon2004tensile}) is appealing because it offers a minimization of losses for long-distance propagation via silica waveguides and optical fibers \cite{Nouchi}. 

Given that Si is the leading material in microelectronics and photonics, the fabrication of non-classical light sources within its crystalline structure would simplify the integration of advanced quantum technologies within modern semiconductor production lines. Lately, color centers in Si have been reconsidered in view of quantum information applications, and their fabrication and emission properties are now brought again under the spotlight. To date, a large variety of optically active defects has been identified \cite{davies1989optical, Khoury_Abbarchi_2022} and some irradiation-induced point defects have even demonstrated single photon behavior, namely the W-center \cite{Baron_Durand_Udvarhelyi_Herzig_Khoury_Pezzagna_Meijer_Robert-Philip_2022}, whose zero-phonon line (ZPL) at 1218 nm falls outside the optical communication bands, the G-center \cite{Baron_Durand_Herzig_Khoury_Pezzagna_Meijer_Robert-Philip_2022}, whose ZPL at 1279 nm belongs to the telecom O-band \cite{Khoury_Abbarchi_2022}, and the T-center that also emits in the O-band at 1326 nm \cite{higginbottom2022optical}. Another notable point defect is the C-center. This is formed when mobile carbon atoms are trapped by interstitial oxygen atoms forming interstitial pairs. This defect complex received less attention than the above-mentioned centers, despite being based on common impurities, and its emission line, located at 1569 nm,  falls within the L-band, making it suitable for long-haul interconnections.
\\Electron irradiation \cite{Chartrand_Bergeron_Morse_Riemann_Abrosimov_Becker_Phol_Simmons_Thewalt_2018, jones1973temperature, Ishikawa_Koga_Itahashi_Vlasenko_Itoh_2009, bohnert1993transient, nakamura1994photoluminescence, kurner1989structure, davies1989temperature} is the most common method employed in literature to generate defects in a Si target. Electrons beams can be easily controlled and deposit damage relatively uniformly over large depths. However, they are less effective than ions at producing lattice displacement damage. The latter offer large stopping power and a well-defined Bragg peak, corresponding to a specific region where the largest amount of defects is localized. This can be suitably varied by controlling the ion's energy, and provides the spatial selectivity that is relevant in many semiconductor applications, particularly in semiconductor heterostructures. Yet, little attention has been paid to characterize the effects of light ions as possible means to implement telecom sources in Si \cite{Davies_Hayama_Murin_2006, Baron_Durand_Herzig_Khoury_Pezzagna_Meijer_Robert-Philip_2022, hallen1996lifetime}. This work explores such possibility reporting the use of $\mathrm{H^+}$ (protons) and $\mathrm{He^+}$ ($\alpha$ particles) with energy of $\sim$ MeV for the generation and characterization of color centers in Si, with particular focus on the C-center. We investigate the temperature dependence of the ZPL fluorescence, proving evidence for the interplay between the various energy levels of the defect. Moreover, time-resolved PL (TRPL) measurements support the close presence of a singlet-allowed transition to dark nonradiative states. Finally, we demonstrate the formation of C-centers through $\mathrm{He^+}$ irradiation of Si-based heterostructures. These epitaxial systems can open novel perspectives for applications such as long-haul secure communications and distributed quantum computing. 
\section{Results and Discussion}\label{sec2}

\subsection{Light-ion generation of telecom emitters in Si}

\begin{figure*} 
\includegraphics[width=\linewidth]{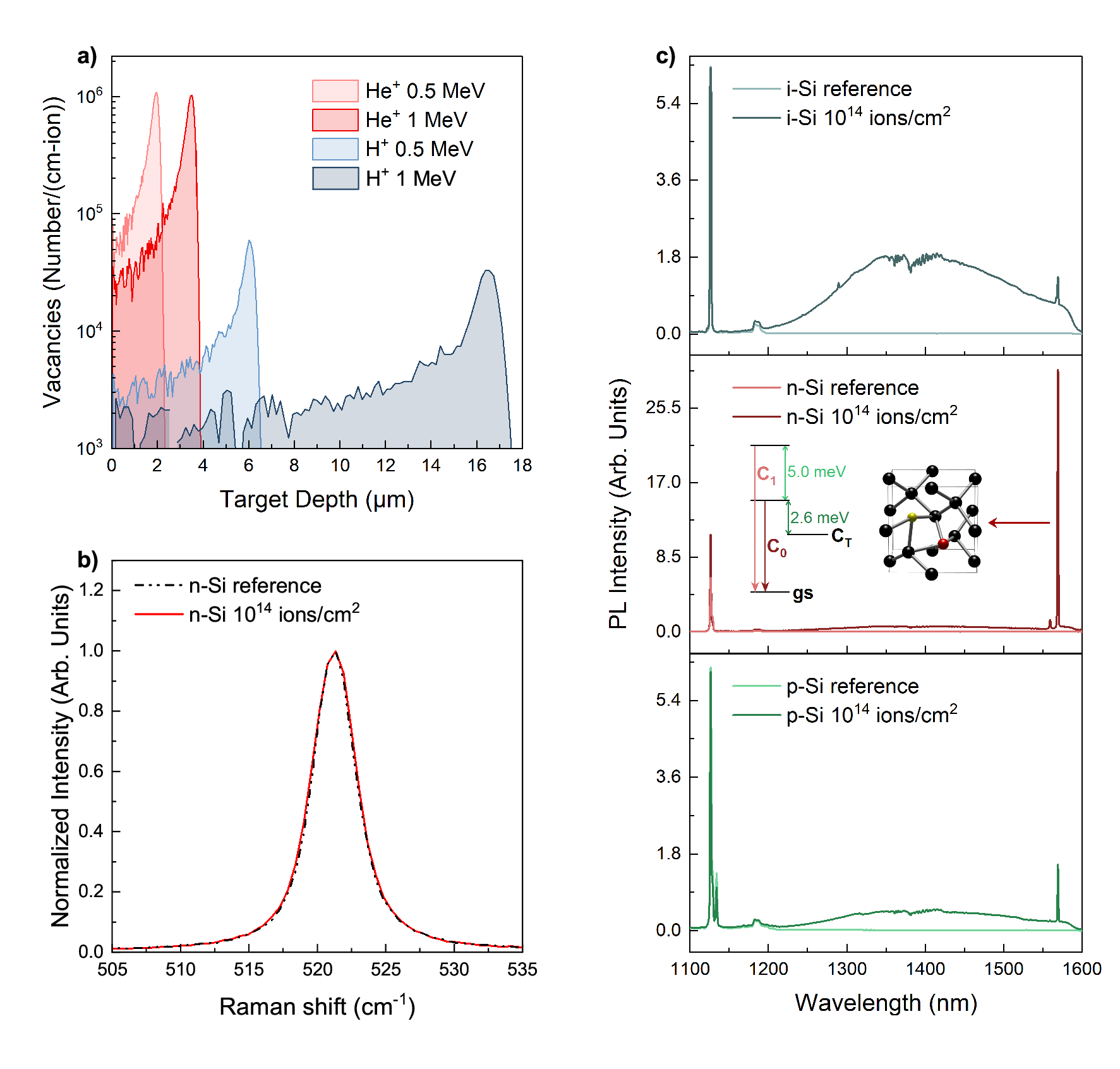}
\caption{\textbf{Effect of $\mathrm{H^+}$ and $\mathrm{He^+}$  irradiation on silicon with different doping types.} \textbf{a)} Calculations based on the Transport of Ions in Matter (TRIM) software of the vacancy distributions generated in Si by $\mathrm{H^+}$ and $\mathrm{He^+}$  with energies of 0.5 and 1 MeV; \textbf{b)} Raman spectra of n-Si (irradiated with $\mathrm{He^+}$  at 0.5 MeV and a dose of 10\textsuperscript{14} ions/cm\textsuperscript{2}) comparing the irradiated and not-irradiated regions; \textbf{c)} PL spectra at 4 K of a not-irradiated Si sample and an irradiated one with 1 MeV $\mathrm{H^+}$ and a dose of 10\textsuperscript{14} ions/cm\textsuperscript{2} for three different doping materials: i-Si, n-Si and p-Si. The middle panel shows the atomic and electronic structures of the C-center.}
\label{Fig1}
\end{figure*}

Before experimentally investigating the irradiation effects on Si, we used the Transport of Ions in Matter (TRIM) software to estimate the vacancy distribution along the target depth for different ion - energy combinations (see Methods) \cite{biersack80, ziegler1999stopping, Agarwal_Lin_Li_Stoller_Zinkle_2021}. The results shown in Fig. \ref{Fig1}a demonstrate that $\mathrm{H^+}$ exhibits a larger penetration depth in Si compared to $\mathrm{He^+}$  at the same energy, owing to the different stopping power associated with their respective masses \cite{ziegler1999stopping}. 

We then performed Raman spectroscopy to use vibrational modes as a proxy for structural damage caused by the ions. Fig. \ref{Fig1}b presents the characteristic one-phonon Raman peak of Si at $\mathrm{520.7 \,cm^{-1}}$ for both the pristine and irradiated regions under a 532 nm excitation. The latter refers to 0.5 MeV $\mathrm{He^+}$  and a dose of 10\textsuperscript{14} ions/cm\textsuperscript{2}. Fig. \ref{Fig1}b reveals no changes in the shape or spectral position of the Raman mode, suggesting that the irradiation process does not affect the long-range atomic order. It gently perturbs the original structural and vibrational properties of the material, while yielding the potential generation of vacancies and thus optically-active point defects as suggested by TRIM calculations. It is worth noting that these Raman results, while limited to the surface because of the sub-$\mu$m penetration depth of light in the visible range, are satisfactorily confirmed by the Raman data obtained by means of an infrared excitation (see Fig. \ref{Fig1}c and the following discussion of the low-temperature investigation).

According to the simulations of the ion transport in matter, $\mathrm{H^+}$ at 1 MeV create vacancies across the largest volume of the target (see Fig. \ref{Fig1}a). This in turn, generates the greatest number of defects among the selected conditions, thus facilitating the optical investigation of the ensemble properties. To explore also the potential influence of doping and growth method on the formation of the irradiation-induced point defects associated to the ubiquitous C and O impurities, we compare intrinsic and doped wafers as Si:P and Si:B, hereafter termed i-Si, n-Si and p-Si, respectively, at a fixed dose of 10\textsuperscript{14} ions/cm\textsuperscript{2} (see Methods). 

The low temperature photoluminescence (PL) spectra are reported in Fig. \ref{Fig1}c. From shorter wavelengths, two peaks are clearly visible in both pristine and irradiated samples, corresponding to the one- and two-phonon Raman spectra centered at 1126 and 1188 nm, respectively \cite{achilli_25}. Additionally, in the p-Si sample, the emission line of the B bound exciton is sufficiently bright to be observed at 1133 nm \cite{tajima1977photoluminescence, PhysRevLett.86.6010}. Notably, the spectra of all the irradiated wafers exhibit a sizable weight between 1150 and 1600 nm. Such broad PL structure is associated with the overall lattice disorder induced by the incident ion beam \cite{Davies_Hayama_Murin_2006}. Particularly, in irradiated i-Si, a sharp peak emerges within the broad band at 1278 nm, corresponding to the well-known emission line of the G-center \cite{Davies_Hayama_Murin_2006, Chartrand_Bergeron_Morse_Riemann_Abrosimov_Becker_Phol_Simmons_Thewalt_2018}. At longer wavelengths, the most prominent PL that spontaneously emerges after irradiation is located at 1569 nm and is ascribed to the emission from the first excited singlet state of the C-center, known as C\textsubscript{0}-line \cite{Wagner_Thonke_Sauer_1984}. Remarkably, this outweighs all the other spectral features. Moreover, in the irradiated n-Si an additional line can be distinctly seen at 1559 nm, which can be ascribed to the transition involving the second non-degenerate level of the first excited singlet state of the C-center, commonly referred to as the C\textsubscript{1}-line \cite{Thonke_Wagner_Hangleiter_Sauer_1985}. We emphasize that the sharp spectral features emerge even without any thermal annealing procedure, thus demonstrating the efficiency of the light ions in generating optically-active C-centers. It is illuminating to notice that neither the broad band nor the sharp peaks of the emission by G- and C-centers are present in the reference samples for any of the doping content, providing a conclusive demonstration that these spectral features are induced by ion irradiation. 

The C-center structure consisting of the (O\textsubscript{i}-C\textsubscript{i}) interstitial pair is illustrated in the inset of Fig. \ref{Fig1}c along with a schematics of its associated electronic levels and optical transitions. Present models suggest that the defect lays in the Si band gap and its ground state is a singlet level that originates from the p-type orbital localized on the C atom. \cite{Silkinis25,udvarhelyi2022band} Under optical pumping, the interaction between such charged defect and a bound carrier can result in a set of bound exciton states. Finally, a triplet state is expect to be $\sim$ 3 meV below the singlet excited state \cite{Silkinis25,udvarhelyi2022band, Ishikawa_Koga_Itahashi_Vlasenko_Itoh_2009, Thonke_Wagner_Hangleiter_Sauer_1985, Wagner_Thonke_Sauer_1984}. The comparison between the irradiated samples clearly highlights that n-Si is more susceptible to irradiation than i-Si and p-Si. The difference in the C-center signal intensity between n-Si and i-Si can be attributed to the respective growth methods: the float zone process used for i-Si introduces significantly less oxygen contamination in the final ingot compared to the Czochralski process used for n-Si and p-Si \cite{LIU201558, LAPPA20041, newman2000oxygen, NAGAI2014737, MULLER2005262}. Consequently, this limits the number of impurities available to form the C-center in i-Si. We speculate that the difference in the C-line intensity between n-Si and p-Si may be related to the formation of B-O complexes in the latter \cite{schutz2010light, schmidt2004structure}. This thermally activated process competes with the C-center generation, thus reducing the number of O atoms available for the formation of C-centers in p-Si. 

\subsection{Fluorescence properties of the C-center}

\begin{figure*}
\includegraphics[width= 12 cm]{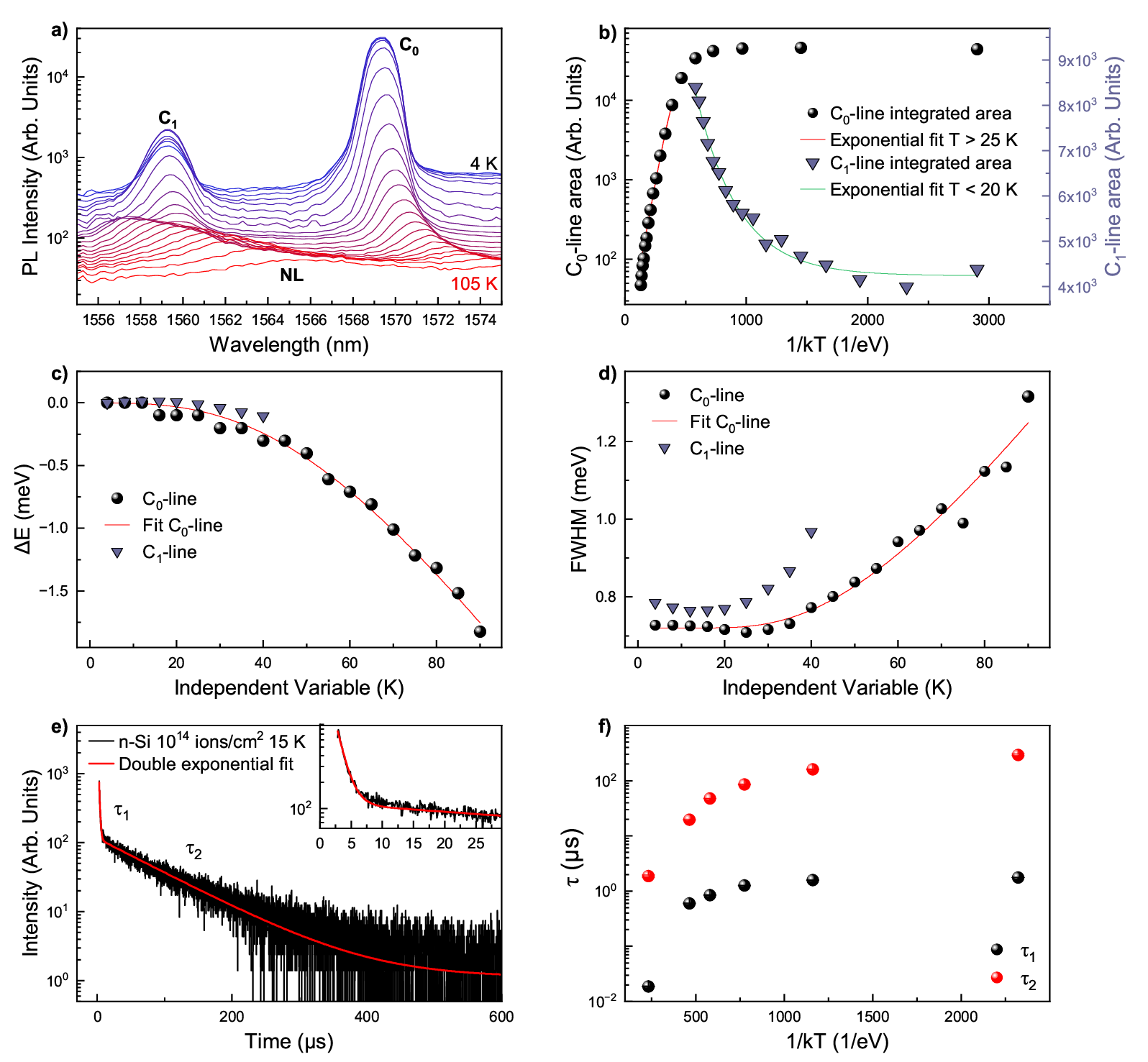}
\caption{\textbf{Characterization of the C-center generated in irradiated n-Si.} \textbf{a)} Ensemble PL spectra showing of the C\textsubscript{0}- and C\textsubscript{1}-lines and a new line, NL, likely not belonging to the C-center, generated by 1 MeV $\mathrm{H^+}$ and a dose of 10\textsuperscript{14} ions/cm\textsuperscript{2}. The spectra are measured from 4 K to 105 K; \textbf{b)} Integrated area of the C\textsubscript{0} peak reported in panel a) as a function of the temperature. The solid line is the fit of the experimental data for T \textgreater \, 25 K with an Arrhenius function (Eq. \ref{eq:T_fit}). Integrated area of the C\textsubscript{1} peaks - generated by 0.5 MeV $\mathrm{He^+}$  irradiation with a dose of 5$\times$10\textsuperscript{14} ions/cm\textsuperscript{2} - and Arrhenius fit (solid line) for \textless \, 20 K; \textbf{c)} Temperature dependence of $\Delta E=E_{ZPL}(T)-E_{ZPL}(T=0 K)$ and associated fit (solid line) for the C\textsubscript{0}-line with the model proposed by Ref. \cite{passler1997basic}; \textbf{d)} Temperature dependence of the full width at half maximum (FWHM) and modeling (solid line) based on Eq. \ref{eq:G_fit}; \textbf{e)} Experimental PL lifetime measured at 15 K of the C\textsubscript{0}-line generated by 1 MeV $\mathrm{H^+}$ and a dose of 10\textsuperscript{14} ions/cm\textsuperscript{2} along with the corresponding double-exponential fit (solid line). The inset shows a magnified view of the curve within the first $\sim$25 $\mu$s. \textbf{f)} Temperature dependence of the two decay components ($\tau_1$ and $\tau_2$) in a logarithmic scale.}
\label{Fig2}
\end{figure*}

In the following, we focus on n-Si to study in detail the fluorescence properties of an ensemble of C-centers. More specifically, we characterize the C-center emission by performing temperature- and lifetime-dependent PL measurements. Fig. \ref{Fig2}a demonstrates that the intensity of the C\textsubscript{0}-line decreases by increasing the lattice temperature up to 100 K. Above this temperature, the peak becomes indistinguishable from the noise. The Arrhenius plot shown in Fig. \ref{Fig2}b for the integrated intensity reveals two different temperature regimes. At low temperatures, i.e., T \textless \, 20 K, the C\textsubscript{0}-line increases. This is likely caused by the thermal depopulation of a dark state, e.g., the triplet level which is expected to be 2.64 meV below the singlet state (see Fig. \ref{Fig1}c) \cite{Chartrand_Bergeron_Morse_Riemann_Abrosimov_Becker_Phol_Simmons_Thewalt_2018, Ishikawa_Koga_Itahashi_Vlasenko_Itoh_2009}. In contrast, when T \textgreater \, 20 K, the C\textsubscript{0}-line is progressively quenched. The experimental data in this high-temperature regime can be fitted using the following equation \cite{Davies_Brian_Lightowlers_Barraclough_THomaz_1989} (red curve in Fig. \ref{Fig2}b):
\begin{equation} \label{eq:T_fit}
    I(T)=\frac{I_0}{1+\rho\mathrm{e}^{-\frac{E_a}{k_BT}}}  
\end{equation}
where $\rho$ is a constant, k\textsubscript{B} is the Boltzmann constant, and E\textsubscript{a} is the activation energy of the process.
\\Such modeling allows us to determine the E\textsubscript{a} pertaining to the non-radiative process responsible for the reduction of emission from the first excited state of the C-center. The average E\textsubscript{a} value, calculated using five samples with varying doses (ranging from $5\times10$\textsuperscript{13} ions/cm\textsuperscript{2} to 10\textsuperscript{16} ions/cm\textsuperscript{2}) in the same irradiation series of 1 MeV $\mathrm{H^+}$, is 23.0 $\pm$ 0.6 meV. This result is consistent with values previously reported in the literature \cite{Streetman_Compton_Johnson_Jones_Noonan_Wong_1971, Jones_Johnson_Compton_Noonan_Streetman_1973}.   

Interestingly, the C\textsubscript{1}-line shows a puzzling behavior increasing from 4 K to 20 K, as reported by the purple triangles in Fig. \ref{Fig2}b. The experimental data follow again an Arrhenius behavior (green curve) with E\textsubscript{a} $=2.9 \pm 0.2$ meV. This value can be interpreted using the present understanding of the electronic structure of the defect, with the C\textsubscript{1} state being thermally populated by the C\textsubscript{0} (see Fig. \ref{Fig1}c). The experimental value derived here compares indeed well with the expected energy distance of 5 meV between the C\textsubscript{1} and the C\textsubscript{0} levels  \cite{Thonke_Wagner_Hangleiter_Sauer_1985}. 

A new line (NL) becomes stronger than the C\textsubscript{1}-line when T \textgreater \, 50 K, as shown in Fig. \ref{Fig2}a. However, the temperature-induced redshift of NL is rather pronounced and markedly different from those of both C\textsubscript{1} and C\textsubscript{0}, thereby suggesting that this emission is likely not related to the C-center. Unveiling its origin and behavior require nevertheless future dedicated studies. The peak energy and full-width at half-maximum (FWHM) of the C-center transitions are shown  in detail as a function of the temperature in Fig. 2c and Fig. 2d, respectively. Up to 40 K, the C\textsubscript{1}-line position and FWHM (purple triangles of Fig. 2c and Fig. 2d) clearly mimic the behavior to the brightest C\textsubscript{0}-line. The variation of the ZPL energy of the latter, $\Delta E=E(T)-E(T=0)$, can be followed over a broader range and nicely follows the conventional dependency of the fundamental energy gaps in semiconductor, which in the analytical representation proposed by P\"assler is defined by the effective phonon temperature $\Theta_p=126 \pm 11$ K and the parameters $\alpha=0.048 \pm 0.005$ meV/K and $p=2.9 \pm 0.5$ \cite{passler1997basic}. The FWHM can be expressed as a function of temperature as \cite{quard2024femtosecond}:
\begin{equation} \label{eq:G_fit}
    \Gamma(T)=\Gamma_o + \frac{a}{\mathrm{e}^{-\frac{\Omega}{k_BT}}-1}  
\end{equation}
providing a zero temperature limit $\Gamma_0=0.72 \pm 0.01$ meV and the intensity and energy of the phonon-emitter coupling: $a=2.7\pm0.8$ meV and $\Omega=14 \pm 2$ meV, respectively. It is worth noting that these values for the C-center are in good agreement with results reported for other quantum emitters in Si, such as the W- and G-centers \cite{quard2024femtosecond}. 

Additional information about the nature and the photophysics pertaining to the electronic states of this notable carbon-oxygen complex can be obtained by investigating its lifetime. As shown in Fig.~\ref{Fig2}e, at low temperature, the TRPL of the C\textsubscript{0}-line demonstrates a well-resolved double-exponential decay with characteristic values of $\tau_1 = (1.264 \pm 0.007) \, \mu\text{s}$ and $\tau_2 = (85.9 \pm 0.4) \, \mu\text{s}$. This particular transient highlights the occurrence of two distinct recombination pathways, whose overall dynamics accelerates when the temperature increases: a behavior proved in Fig.~\ref{Fig2}f by a substantial decrease of the lifetime between 5 and 50 K. More importantly, we have found that as the temperature increases up to 25 K, the amplitude of the second component $\tau_2$ gains dominance over the counterpart $\tau_1$. Such findings are fully consistent with the results reported by Thonke et al.~\cite{Thonke_Wagner_Hangleiter_Sauer_1985} and suggest that the C\textsubscript{0}-line can be interpreted as a singlet transition that is inherently characterized by a relatively fast ($\tau_1$) lifetime, while the long-lived ($\tau_2$) component stems from the temperature-induced population caused by a lower-lying dark level in the form of a spin-triplet state bound to the same defect \cite{Thonke_Wagner_Hangleiter_Sauer_1985, udvarhelyi2022band}.

\subsection{Irradiation of Ge-on-Si heterostructures}

\begin{figure*}
\includegraphics[width=\linewidth]{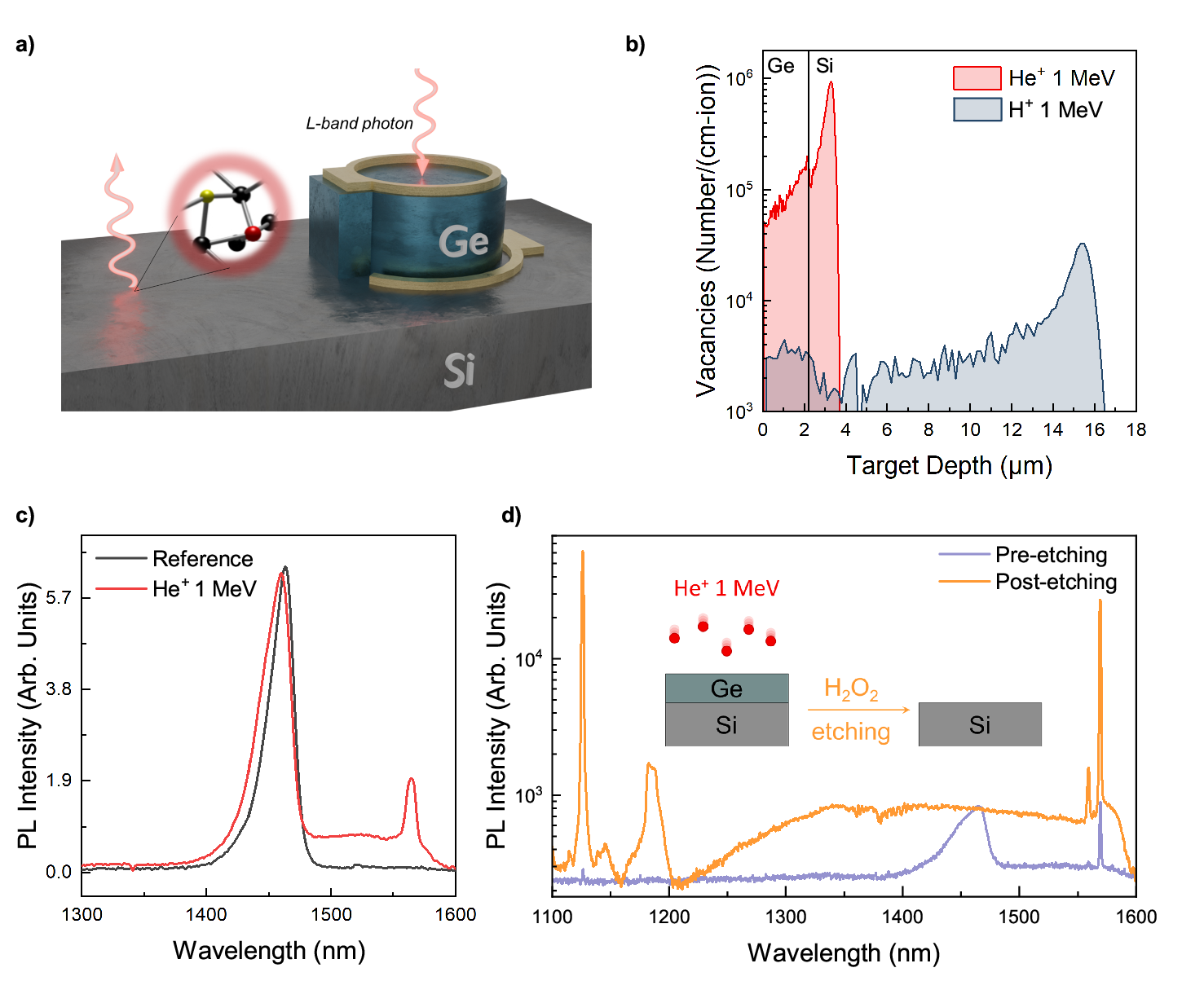}
\caption{\textbf{Generation of C-centers in Ge-on-Si heterostructures.} \textbf{a)} Artistic concept of a Ge-on-Si-based quantum device embedding emitters in Si in the form of C-centers, while the photodetection functionality is provided by Ge. \textbf{b)} Vacancy distributions computed by TRIM for 1 MeV $\mathrm{He^+}$  and $\mathrm{H^+}$ in Ge-on-Si heterostructure; \textbf{c)} Low temperature PL spectra of a non-irradiated Ge-on-Si heterostructure (grey curve) and an irradiated one with $\mathrm{He^+}$  of 1 MeV and a dose of 10\textsuperscript{15} ions/cm\textsuperscript{2} (red curve); \textbf{d)} Spectra of the irradiated Ge-on-Si sample with $\mathrm{He^+}$ of 1 MeV and a dose of 10\textsuperscript{15} ions/cm\textsuperscript{2} (purple curve) and of the same heterostructure after the removal of the Ge top layer through H\textsubscript{2}O\textsubscript{2} etching (yellow curve).}
\label{Fig3}
\end{figure*}

After demonstrating the potential of light ions to yield the formation of C-centers in Si, we can explore their viable use in innovative and advanced epitaxial systems that can be relevant for future quantum information and communication processing. Here we anticipate Ge-on-Si heterostructures that, thanks to their mature technology, hold the noticeable potential to enable a quantum architecture in which the Si bottom layer contains telecom quantum emitters (C-centers), while the Ge top layer serves as the active medium for single-photon detection \cite{vines_2019, na_2024} (see the conceptual representation in Fig. \ref{Fig3}a). To this end, the optimal configuration would consist of radiation-induced defects that are generated close to the buried Si/Ge interface. Although the C-center does not directly involve the presence of vacancies, as previously shown for the bulk Si case, we assume that a higher concentration of vacancies enhances the atomic mobility, thus increasing the probability to form the C-O complex, i.e., the defect structure that is optically active in the L-band. Fig. \ref{Fig3}b presents the calculated vacancy distributions within the heterostructure for 1 MeV $\mathrm{He^+}$  and $\mathrm{H^+}$. Based on these distributions, we select 1 MeV $\mathrm{He^+}$  to position the Bragg peak at a depth that is a few $\mu$m below the Ge/Si interface. 

To validate the practical generation of C-centers, we compared the PL spectrum of the heterostructure prior to and after irradiation. Fig. \ref{Fig3}c shows that a well-resolved PL peak at $\sim$1450 nm dominates not only the pristine sample, but also the irradiated structure. This spectral feature corresponds to the direct-gap transitions occurring in the Ge epilayer, and suggests, at a first glance, that the radiation parameters utilized in this work guarantee that Ge retains a suitable optical grade despite the lack of any post-irradiation annealing. This is in agreement with the observations on bulk Si reported in the previous sections. More specifically, Fig. \ref{Fig3}c demonstrates the successful generation of C-centers directly in the Si substrate. Indeed, the moderation of $\mathrm{He^+}$  through the epitaxial structure leads to an additional PL peak exactly in the spectral region of the C-line, i.e., at $\sim 1560$ nm. Such feature, shown in Fig. \ref{Fig3}c, is likely due to the convolution of both the C\textsubscript{0} and C\textsubscript{1} lines because of the coarse spectral scan used in this specific measurement. 

A decisive confirmation is provided by Fig. \ref{Fig3}d, which shows higher resolution data over a larger spectral window (purple line). Besides the intense emission from the direct band-gap of Ge, two features ascribed to the Si substrate are chiefly present, namely (i) a weak structure at $\sim1130$ nm, which corresponds to the one-phonon Raman spectrum of Si and (ii) the distinct C\textsubscript{0}-C\textsubscript{1} doublet at $\sim$1570 nm \cite{achilli_25}. We can find reassurance of this by performing a selective wet etching of the Ge thin film by means of H\textsubscript{2}O\textsubscript{2} \cite{huygens2007etching}. Fig. \ref{Fig3}d shows that after chemical dissolution of the Ge layer, the direct transition fully disappears and is substituted by the typical broadened emission related to radiation damage in Si (see, e.g., Fig. \ref{Fig1}). Also, at shorter wavelengths, i.e., between 1100 and 1200 nm, the usually weak second-order phonon Raman scattering emerges and accompanies the now stronger one-phonon Raman mode. Eventually, this provides a compelling demonstration of the genuine excitation of the irradiated volume of the Si substrate. 

Above all, the C\textsubscript{0}-line and C\textsubscript{1}-line intensities drastically increase after wet etching, further confirming that in the heterostructure such excitonic luminescence arises from the effective generation of the C-O complexes in Si owning to the passage of the $\mathrm{He^+}$ ions through the Ge/Si interface.          

\section{Conclusion}\label{sec13}
We have shown that the C-center can be generated in Si by light ion irradiation. The analysis of the defect fluorescence in bulk Si allowed us to select irradiated CZ-grown Si:P as a suitable host. Temperature dependent PL measurements revealed that for T \textless \, 20 K the C\textsubscript{0}-line intensity increases due to the thermal depopulation of the triplet state, whereas for T \textgreater \, 25 K the C\textsubscript{0}-line intensity decreases exponentially with an activation energy of 23.0 ± 0.6 meV. The opposite holds fo the C\textsubscript{1}-line, whose intensity increases in the 4 - 20 K range following an Arrhenius behavior with an activation energy of 2.9 ± 0.2 meV. TRPL revealed a double-exponential transient which is consistent with the thermal depopulation of a lower-lying dark state.  
Finally, we explored ion-irradiation strategies to prove the formation of C-centers in Ge-on-Si heterostructures, which offer novel routes for implementing integrated quantum relays and memories for low-loss long-haul communication and information processing. Through meticulous irradiation experiments, we therefore aim to deepen our understanding of Si color centers and by promoting their application in advanced epitaxial architectures we uncover viable opportunities for furthering the exploitation of quantum photonic concepts.

\section{Methods}\label{sec11}
\textbf{Sample structure and details} The investigated samples consisted of Si wafers and Ge-on-Si heterostructures. The Si samples were commercially available and included (i) a Czochralski (CZ) wafer doped with phosphorous and a resistivity of 1 - 10 $\Omega$cm (n-Si), (ii) a CZ wafer doped with boron (B) and a resistivity of 1 - 10 $\Omega$cm (p-Si), and (iii) an intrinsic float zone (FZ) wafer with a resistivity of 2100 - 3300 $\Omega$cm (i-Si). The Ge-on-Si heterostructures consisted of a CZ wafer onto which a 2.20 $\mu$m thick Ge layer was epitaxially grown by chemical vapor deposition (CVD). 

\textbf{Irradiation protocols} The irradiation procedure was carried out at room temperature using the AN-2000 particle accelerator at the National Institute of Nuclear Physics (INFN) facility in Legnaro (Italy), with $\mathrm{H^+}$ and $\mathrm{He^+}$  having energies of 0.5 or 1 MeV and doses in the 10\textsuperscript{13} - 10\textsuperscript{16} ions/cm\textsuperscript{2} range. The irradiation spot had a dimension of $\sim$4$\times$4 mm\textsuperscript{2}, ensuring that each sample contained a central damaged region surrounded by a pristine frame. 

\textbf{Simulation} TRIM software was used to simulate the damage generated in the target material by the incident ion beam. Specifically, the D1 method proposed by Agarwal et al. \cite{Agarwal_Lin_Li_Stoller_Zinkle_2021} was employed to process the TRIM output files and compute the vacancy distribution more accurately. 

\textbf{Optical investigation} Raman spectroscopy was carried out in a back-scattering configuration using a 532 nm continuous-wave (CW) laser with a power at the samples surface of $\sim$ mW that does not cause illumination-induced heating. The microscope used for the experiment was equipped with a 50$\times$ objective with a numerical aperture of 0.7. Measurements were taken both in the irradiated and in the adjacent non-irradiated regions of the same sample. Fluorescence measurements were conducted in the spectral range between 1100 and 1600 nm, with the samples mounted in a closed-circle cryostat. The excitation source was provided by a CW Nd:YVO$_4$ laser at 1064 nm wth an incident power of 50 mW. The PL was detected with a liquid nitrogen-cooled (In,Ga)As array coupled to a spectrograph.

Further characterization was conducted through lifetime measurements using a 1064 nm pulsed Nd:YVO$_4$ laser operating at a repetition rate of $\mathrm{1\,KHz}$. A NIR 50$\times$ objective with a numerical aperture of 0.42 was used both to focus the laser on the sample and to collect the fluorescence. The PL was filtered by a spectrograph and transmitted via an optical fiber to a Superconducting Nanowire Single-Photon Detector. TRPL was performed using time-correlated single photon counting with a bin width of $\mathrm{100\,ns}$ and an acquisition time of $1200 \,s$. Decay curves were measured as a function of temperature, ranging from 4 K to 50 K, using a closed-cycle cryostat.

\section{acknowledgments}
Financial support was provided by the European Union’s Horizon Europe Research and Innovation Programme under agreement 101070700. Support from PNRR MUR project PE0000023-NQSTI is also acknowledged.


\begin{thebibliography}{43}%
\makeatletter
\providecommand \@ifxundefined [1]{%
 \@ifx{#1\undefined}
}%
\providecommand \@ifnum [1]{%
 \ifnum #1\expandafter \@firstoftwo
 \else \expandafter \@secondoftwo
 \fi
}%
\providecommand \@ifx [1]{%
 \ifx #1\expandafter \@firstoftwo
 \else \expandafter \@secondoftwo
 \fi
}%
\providecommand \natexlab [1]{#1}%
\providecommand \enquote  [1]{``#1''}%
\providecommand \bibnamefont  [1]{#1}%
\providecommand \bibfnamefont [1]{#1}%
\providecommand \citenamefont [1]{#1}%
\providecommand \href@noop [0]{\@secondoftwo}%
\providecommand \href [0]{\begingroup \@sanitize@url \@href}%
\providecommand \@href[1]{\@@startlink{#1}\@@href}%
\providecommand \@@href[1]{\endgroup#1\@@endlink}%
\providecommand \@sanitize@url [0]{\catcode `\\12\catcode `\$12\catcode `\&12\catcode `\#12\catcode `\^12\catcode `\_12\catcode `\%12\relax}%
\providecommand \@@startlink[1]{}%
\providecommand \@@endlink[0]{}%
\providecommand \url  [0]{\begingroup\@sanitize@url \@url }%
\providecommand \@url [1]{\endgroup\@href {#1}{\urlprefix }}%
\providecommand \urlprefix  [0]{URL }%
\providecommand \Eprint [0]{\href }%
\providecommand \doibase [0]{https://doi.org/}%
\providecommand \selectlanguage [0]{\@gobble}%
\providecommand \bibinfo  [0]{\@secondoftwo}%
\providecommand \bibfield  [0]{\@secondoftwo}%
\providecommand \translation [1]{[#1]}%
\providecommand \BibitemOpen [0]{}%
\providecommand \bibitemStop [0]{}%
\providecommand \bibitemNoStop [0]{.\EOS\space}%
\providecommand \EOS [0]{\spacefactor3000\relax}%
\providecommand \BibitemShut  [1]{\csname bibitem#1\endcsname}%
\let\auto@bib@innerbib\@empty
\bibitem [{\citenamefont {Andrini}\ \emph {et~al.}(2024)\citenamefont {Andrini}, \citenamefont {Amanti}, \citenamefont {Armani}, \citenamefont {Bellani}, \citenamefont {Bonaiuto}, \citenamefont {Cammarata}, \citenamefont {Campostrini}, \citenamefont {Dao}, \citenamefont {De~Matteis}, \citenamefont {Demontis} \emph {et~al.}}]{andrini2024solid}%
  \BibitemOpen
  \bibfield  {author} {\bibinfo {author} {\bibfnamefont {G.}~\bibnamefont {Andrini}}, \bibinfo {author} {\bibfnamefont {F.}~\bibnamefont {Amanti}}, \bibinfo {author} {\bibfnamefont {F.}~\bibnamefont {Armani}}, \bibinfo {author} {\bibfnamefont {V.}~\bibnamefont {Bellani}}, \bibinfo {author} {\bibfnamefont {V.}~\bibnamefont {Bonaiuto}}, \bibinfo {author} {\bibfnamefont {S.}~\bibnamefont {Cammarata}}, \bibinfo {author} {\bibfnamefont {M.}~\bibnamefont {Campostrini}}, \bibinfo {author} {\bibfnamefont {T.~H.}\ \bibnamefont {Dao}}, \bibinfo {author} {\bibfnamefont {F.}~\bibnamefont {De~Matteis}}, \bibinfo {author} {\bibfnamefont {V.}~\bibnamefont {Demontis}}, \emph {et~al.},\ }\bibfield  {title} {\bibinfo {title} {Solid-state color centers for single-photon generation},\ }in\ \href@noop {} {\emph {\bibinfo {booktitle} {Photonics}}},\ Vol.~\bibinfo {volume} {11}\ (\bibinfo {organization} {MDPI},\ \bibinfo {year} {2024})\ p.\ \bibinfo {pages} {188}\BibitemShut {NoStop}%
\bibitem [{\citenamefont {Aberl}\ \emph {et~al.}(2024)\citenamefont {Aberl}, \citenamefont {Navarrete}, \citenamefont {Karaman}, \citenamefont {Enriquez}, \citenamefont {Wilflingseder}, \citenamefont {Salomon}, \citenamefont {Primetzhofer}, \citenamefont {Schubert}, \citenamefont {Capellini}, \citenamefont {Fromherz} \emph {et~al.}}]{aberl2024all}%
  \BibitemOpen
  \bibfield  {author} {\bibinfo {author} {\bibfnamefont {J.}~\bibnamefont {Aberl}}, \bibinfo {author} {\bibfnamefont {E.~P.}\ \bibnamefont {Navarrete}}, \bibinfo {author} {\bibfnamefont {M.}~\bibnamefont {Karaman}}, \bibinfo {author} {\bibfnamefont {D.~H.}\ \bibnamefont {Enriquez}}, \bibinfo {author} {\bibfnamefont {C.}~\bibnamefont {Wilflingseder}}, \bibinfo {author} {\bibfnamefont {A.}~\bibnamefont {Salomon}}, \bibinfo {author} {\bibfnamefont {D.}~\bibnamefont {Primetzhofer}}, \bibinfo {author} {\bibfnamefont {M.~A.}\ \bibnamefont {Schubert}}, \bibinfo {author} {\bibfnamefont {G.}~\bibnamefont {Capellini}}, \bibinfo {author} {\bibfnamefont {T.}~\bibnamefont {Fromherz}}, \emph {et~al.},\ }\bibfield  {title} {\bibinfo {title} {All-epitaxial self-assembly of silicon color centers confined within sub-nanometer thin layers using ultra-low temperature epitaxy},\ }\href@noop {} {\bibfield  {journal} {\bibinfo  {journal} {Advanced Materials}\ }\textbf {\bibinfo {volume} {36}},\ \bibinfo {pages} {2408424} (\bibinfo
  {year} {2024})}\BibitemShut {NoStop}%
\bibitem [{\citenamefont {Cannon}\ \emph {et~al.}(2004)\citenamefont {Cannon}, \citenamefont {Liu}, \citenamefont {Ishikawa}, \citenamefont {Wada}, \citenamefont {Danielson}, \citenamefont {Jongthammanurak}, \citenamefont {Michel},\ and\ \citenamefont {Kimerling}}]{cannon2004tensile}%
  \BibitemOpen
  \bibfield  {author} {\bibinfo {author} {\bibfnamefont {D.~D.}\ \bibnamefont {Cannon}}, \bibinfo {author} {\bibfnamefont {J.}~\bibnamefont {Liu}}, \bibinfo {author} {\bibfnamefont {Y.}~\bibnamefont {Ishikawa}}, \bibinfo {author} {\bibfnamefont {K.}~\bibnamefont {Wada}}, \bibinfo {author} {\bibfnamefont {D.~T.}\ \bibnamefont {Danielson}}, \bibinfo {author} {\bibfnamefont {S.}~\bibnamefont {Jongthammanurak}}, \bibinfo {author} {\bibfnamefont {J.}~\bibnamefont {Michel}},\ and\ \bibinfo {author} {\bibfnamefont {L.~C.}\ \bibnamefont {Kimerling}},\ }\bibfield  {title} {\bibinfo {title} {Tensile strained epitaxial ge films on si (100) substrates with potential application in l-band telecommunications},\ }\href@noop {} {\bibfield  {journal} {\bibinfo  {journal} {Applied Physics Letters}\ }\textbf {\bibinfo {volume} {84}},\ \bibinfo {pages} {906} (\bibinfo {year} {2004})}\BibitemShut {NoStop}%
\bibitem [{\citenamefont {Nouchi}\ \emph {et~al.}(2003)\citenamefont {Nouchi}, \citenamefont {Montmorillon}, \citenamefont {Sillard}, \citenamefont {Bertaina},\ and\ \citenamefont {Guenot}}]{Nouchi}%
  \BibitemOpen
  \bibfield  {author} {\bibinfo {author} {\bibfnamefont {P.}~\bibnamefont {Nouchi}}, \bibinfo {author} {\bibfnamefont {L.-A.}\ \bibnamefont {Montmorillon}}, \bibinfo {author} {\bibfnamefont {P.}~\bibnamefont {Sillard}}, \bibinfo {author} {\bibfnamefont {A.}~\bibnamefont {Bertaina}},\ and\ \bibinfo {author} {\bibfnamefont {P.}~\bibnamefont {Guenot}},\ }\bibfield  {title} {\bibinfo {title} {Optical fiber design for wavelength-multiplexed transmission - conception des fibres optiques pour la transmission multiplexée en longueur d'onde},\ }\href@noop {} {\bibfield  {journal} {\bibinfo  {journal} {C. R. Physique}\ }\textbf {\bibinfo {volume} {4}},\ \bibinfo {pages} {29} (\bibinfo {year} {2003})}\BibitemShut {NoStop}%
\bibitem [{\citenamefont {Davies}(1989)}]{davies1989optical}%
  \BibitemOpen
  \bibfield  {author} {\bibinfo {author} {\bibfnamefont {G.}~\bibnamefont {Davies}},\ }\bibfield  {title} {\bibinfo {title} {The optical properties of luminescence centres in silicon},\ }\href@noop {} {\bibfield  {journal} {\bibinfo  {journal} {Physics reports}\ }\textbf {\bibinfo {volume} {176}},\ \bibinfo {pages} {83} (\bibinfo {year} {1989})}\BibitemShut {NoStop}%
\bibitem [{\citenamefont {Khoury}\ and\ \citenamefont {Abbarchi}(2022)}]{Khoury_Abbarchi_2022}%
  \BibitemOpen
  \bibfield  {author} {\bibinfo {author} {\bibfnamefont {M.}~\bibnamefont {Khoury}}\ and\ \bibinfo {author} {\bibfnamefont {M.}~\bibnamefont {Abbarchi}},\ }\bibfield  {title} {\bibinfo {title} {A bright future for silicon in quantum technologies},\ }\href@noop {} {\bibfield  {journal} {\bibinfo  {journal} {Journal of Applied Physics}\ }\textbf {\bibinfo {volume} {131}} (\bibinfo {year} {2022})}\BibitemShut {NoStop}%
\bibitem [{\citenamefont {Baron}\ \emph {et~al.}(2022{\natexlab{a}})\citenamefont {Baron}, \citenamefont {Durand}, \citenamefont {Udvarhelyi}, \citenamefont {Herzig}, \citenamefont {Khoury}, \citenamefont {Pezzagna}, \citenamefont {Meijer}, \citenamefont {Robert-Philip}, \citenamefont {Abbarchi}, \citenamefont {Hartmann} \emph {et~al.}}]{Baron_Durand_Udvarhelyi_Herzig_Khoury_Pezzagna_Meijer_Robert-Philip_2022}%
  \BibitemOpen
  \bibfield  {author} {\bibinfo {author} {\bibfnamefont {Y.}~\bibnamefont {Baron}}, \bibinfo {author} {\bibfnamefont {A.}~\bibnamefont {Durand}}, \bibinfo {author} {\bibfnamefont {P.}~\bibnamefont {Udvarhelyi}}, \bibinfo {author} {\bibfnamefont {T.}~\bibnamefont {Herzig}}, \bibinfo {author} {\bibfnamefont {M.}~\bibnamefont {Khoury}}, \bibinfo {author} {\bibfnamefont {S.}~\bibnamefont {Pezzagna}}, \bibinfo {author} {\bibfnamefont {J.}~\bibnamefont {Meijer}}, \bibinfo {author} {\bibfnamefont {I.}~\bibnamefont {Robert-Philip}}, \bibinfo {author} {\bibfnamefont {M.}~\bibnamefont {Abbarchi}}, \bibinfo {author} {\bibfnamefont {J.-M.}\ \bibnamefont {Hartmann}}, \emph {et~al.},\ }\bibfield  {title} {\bibinfo {title} {Detection of single w-centers in silicon},\ }\href@noop {} {\bibfield  {journal} {\bibinfo  {journal} {ACS photonics}\ }\textbf {\bibinfo {volume} {9}},\ \bibinfo {pages} {2337} (\bibinfo {year} {2022}{\natexlab{a}})}\BibitemShut {NoStop}%
\bibitem [{\citenamefont {Baron}\ \emph {et~al.}(2022{\natexlab{b}})\citenamefont {Baron}, \citenamefont {Durand}, \citenamefont {Herzig}, \citenamefont {Khoury}, \citenamefont {Pezzagna}, \citenamefont {Meijer}, \citenamefont {Robert-Philip}, \citenamefont {Abbarchi}, \citenamefont {Hartmann}, \citenamefont {Reboh} \emph {et~al.}}]{Baron_Durand_Herzig_Khoury_Pezzagna_Meijer_Robert-Philip_2022}%
  \BibitemOpen
  \bibfield  {author} {\bibinfo {author} {\bibfnamefont {Y.}~\bibnamefont {Baron}}, \bibinfo {author} {\bibfnamefont {A.}~\bibnamefont {Durand}}, \bibinfo {author} {\bibfnamefont {T.}~\bibnamefont {Herzig}}, \bibinfo {author} {\bibfnamefont {M.}~\bibnamefont {Khoury}}, \bibinfo {author} {\bibfnamefont {S.}~\bibnamefont {Pezzagna}}, \bibinfo {author} {\bibfnamefont {J.}~\bibnamefont {Meijer}}, \bibinfo {author} {\bibfnamefont {I.}~\bibnamefont {Robert-Philip}}, \bibinfo {author} {\bibfnamefont {M.}~\bibnamefont {Abbarchi}}, \bibinfo {author} {\bibfnamefont {J.-M.}\ \bibnamefont {Hartmann}}, \bibinfo {author} {\bibfnamefont {S.}~\bibnamefont {Reboh}}, \emph {et~al.},\ }\bibfield  {title} {\bibinfo {title} {Single g centers in silicon fabricated by co-implantation with carbon and proton},\ }\href@noop {} {\bibfield  {journal} {\bibinfo  {journal} {Applied Physics Letters}\ }\textbf {\bibinfo {volume} {121}} (\bibinfo {year} {2022}{\natexlab{b}})}\BibitemShut {NoStop}%
\bibitem [{\citenamefont {Higginbottom}\ \emph {et~al.}(2022)\citenamefont {Higginbottom}, \citenamefont {Kurkjian}, \citenamefont {Chartrand}, \citenamefont {Kazemi}, \citenamefont {Brunelle}, \citenamefont {MacQuarrie}, \citenamefont {Klein}, \citenamefont {Lee-Hone}, \citenamefont {Stacho}, \citenamefont {Ruether} \emph {et~al.}}]{higginbottom2022optical}%
  \BibitemOpen
  \bibfield  {author} {\bibinfo {author} {\bibfnamefont {D.~B.}\ \bibnamefont {Higginbottom}}, \bibinfo {author} {\bibfnamefont {A.~T.}\ \bibnamefont {Kurkjian}}, \bibinfo {author} {\bibfnamefont {C.}~\bibnamefont {Chartrand}}, \bibinfo {author} {\bibfnamefont {M.}~\bibnamefont {Kazemi}}, \bibinfo {author} {\bibfnamefont {N.~A.}\ \bibnamefont {Brunelle}}, \bibinfo {author} {\bibfnamefont {E.~R.}\ \bibnamefont {MacQuarrie}}, \bibinfo {author} {\bibfnamefont {J.~R.}\ \bibnamefont {Klein}}, \bibinfo {author} {\bibfnamefont {N.~R.}\ \bibnamefont {Lee-Hone}}, \bibinfo {author} {\bibfnamefont {J.}~\bibnamefont {Stacho}}, \bibinfo {author} {\bibfnamefont {M.}~\bibnamefont {Ruether}}, \emph {et~al.},\ }\bibfield  {title} {\bibinfo {title} {Optical observation of single spins in silicon},\ }\href@noop {} {\bibfield  {journal} {\bibinfo  {journal} {Nature}\ }\textbf {\bibinfo {volume} {607}},\ \bibinfo {pages} {266} (\bibinfo {year} {2022})}\BibitemShut {NoStop}%
\bibitem [{\citenamefont {Chartrand}\ \emph {et~al.}(2018)\citenamefont {Chartrand}, \citenamefont {Bergeron}, \citenamefont {Morse}, \citenamefont {Riemann}, \citenamefont {Abrosimov}, \citenamefont {Becker}, \citenamefont {Pohl}, \citenamefont {Simmons},\ and\ \citenamefont {Thewalt}}]{Chartrand_Bergeron_Morse_Riemann_Abrosimov_Becker_Phol_Simmons_Thewalt_2018}%
  \BibitemOpen
  \bibfield  {author} {\bibinfo {author} {\bibfnamefont {C.}~\bibnamefont {Chartrand}}, \bibinfo {author} {\bibfnamefont {L.}~\bibnamefont {Bergeron}}, \bibinfo {author} {\bibfnamefont {K.}~\bibnamefont {Morse}}, \bibinfo {author} {\bibfnamefont {H.}~\bibnamefont {Riemann}}, \bibinfo {author} {\bibfnamefont {N.}~\bibnamefont {Abrosimov}}, \bibinfo {author} {\bibfnamefont {P.}~\bibnamefont {Becker}}, \bibinfo {author} {\bibfnamefont {H.-J.}\ \bibnamefont {Pohl}}, \bibinfo {author} {\bibfnamefont {S.}~\bibnamefont {Simmons}},\ and\ \bibinfo {author} {\bibfnamefont {M.}~\bibnamefont {Thewalt}},\ }\bibfield  {title} {\bibinfo {title} {Highly enriched si 28 reveals remarkable optical linewidths and fine structure for well-known damage centers},\ }\href@noop {} {\bibfield  {journal} {\bibinfo  {journal} {Physical Review B}\ }\textbf {\bibinfo {volume} {98}},\ \bibinfo {pages} {195201} (\bibinfo {year} {2018})}\BibitemShut {NoStop}%
\bibitem [{\citenamefont {Jones}\ \emph {et~al.}(1973{\natexlab{a}})\citenamefont {Jones}, \citenamefont {Johnson}, \citenamefont {Compton}, \citenamefont {Noonan},\ and\ \citenamefont {Streetman}}]{jones1973temperature}%
  \BibitemOpen
  \bibfield  {author} {\bibinfo {author} {\bibfnamefont {C.~E.}\ \bibnamefont {Jones}}, \bibinfo {author} {\bibfnamefont {E.~S.}\ \bibnamefont {Johnson}}, \bibinfo {author} {\bibfnamefont {W.~D.}\ \bibnamefont {Compton}}, \bibinfo {author} {\bibfnamefont {J.}~\bibnamefont {Noonan}},\ and\ \bibinfo {author} {\bibfnamefont {B.}~\bibnamefont {Streetman}},\ }\bibfield  {title} {\bibinfo {title} {Temperature, stress, and annealing effects on the luminescence from electron-irradiated silicon},\ }\href@noop {} {\bibfield  {journal} {\bibinfo  {journal} {Journal of Applied Physics}\ }\textbf {\bibinfo {volume} {44}},\ \bibinfo {pages} {5402} (\bibinfo {year} {1973}{\natexlab{a}})}\BibitemShut {NoStop}%
\bibitem [{\citenamefont {Ishikawa}\ \emph {et~al.}(2009)\citenamefont {Ishikawa}, \citenamefont {Koga}, \citenamefont {Itahashi}, \citenamefont {Vlasenko},\ and\ \citenamefont {Itoh}}]{Ishikawa_Koga_Itahashi_Vlasenko_Itoh_2009}%
  \BibitemOpen
  \bibfield  {author} {\bibinfo {author} {\bibfnamefont {T.}~\bibnamefont {Ishikawa}}, \bibinfo {author} {\bibfnamefont {K.}~\bibnamefont {Koga}}, \bibinfo {author} {\bibfnamefont {T.}~\bibnamefont {Itahashi}}, \bibinfo {author} {\bibfnamefont {L.~S.}\ \bibnamefont {Vlasenko}},\ and\ \bibinfo {author} {\bibfnamefont {K.~M.}\ \bibnamefont {Itoh}},\ }\bibfield  {title} {\bibinfo {title} {Photoluminescence from triplet states of isoelectronic bound excitons at carbon-interstitial oxygen defects in silicon},\ }\href@noop {} {\bibfield  {journal} {\bibinfo  {journal} {Physica B}\ }\textbf {\bibinfo {volume} {404}},\ \bibinfo {pages} {4552–4554} (\bibinfo {year} {2009})}\BibitemShut {NoStop}%
\bibitem [{\citenamefont {Bohnert}\ \emph {et~al.}(1993)\citenamefont {Bohnert}, \citenamefont {Weronek},\ and\ \citenamefont {Hangleiter}}]{bohnert1993transient}%
  \BibitemOpen
  \bibfield  {author} {\bibinfo {author} {\bibfnamefont {G.}~\bibnamefont {Bohnert}}, \bibinfo {author} {\bibfnamefont {K.}~\bibnamefont {Weronek}},\ and\ \bibinfo {author} {\bibfnamefont {A.}~\bibnamefont {Hangleiter}},\ }\bibfield  {title} {\bibinfo {title} {Transient characteristics of isoelectronic bound excitons at hole-attractive defects in silicon: The c (0.79 ev), p (0.767 ev), and h (0.926 ev) lines},\ }\href@noop {} {\bibfield  {journal} {\bibinfo  {journal} {Physical Review B}\ }\textbf {\bibinfo {volume} {48}},\ \bibinfo {pages} {14973} (\bibinfo {year} {1993})}\BibitemShut {NoStop}%
\bibitem [{\citenamefont {Nakamura}\ \emph {et~al.}(1994)\citenamefont {Nakamura}, \citenamefont {Kitamura}, \citenamefont {Misawa}, \citenamefont {Suzuki}, \citenamefont {Nagai},\ and\ \citenamefont {Sunaga}}]{nakamura1994photoluminescence}%
  \BibitemOpen
  \bibfield  {author} {\bibinfo {author} {\bibfnamefont {M.}~\bibnamefont {Nakamura}}, \bibinfo {author} {\bibfnamefont {E.}~\bibnamefont {Kitamura}}, \bibinfo {author} {\bibfnamefont {Y.}~\bibnamefont {Misawa}}, \bibinfo {author} {\bibfnamefont {T.}~\bibnamefont {Suzuki}}, \bibinfo {author} {\bibfnamefont {S.}~\bibnamefont {Nagai}},\ and\ \bibinfo {author} {\bibfnamefont {H.}~\bibnamefont {Sunaga}},\ }\bibfield  {title} {\bibinfo {title} {Photoluminescence measurement of carbon in silicon crystals irradiated with high energy electrons},\ }\href@noop {} {\bibfield  {journal} {\bibinfo  {journal} {Journal of the Electrochemical Society}\ }\textbf {\bibinfo {volume} {141}},\ \bibinfo {pages} {3576} (\bibinfo {year} {1994})}\BibitemShut {NoStop}%
\bibitem [{\citenamefont {K{\"u}rner}\ \emph {et~al.}(1989)\citenamefont {K{\"u}rner}, \citenamefont {Sauer}, \citenamefont {D{\"o}rnen},\ and\ \citenamefont {Thonke}}]{kurner1989structure}%
  \BibitemOpen
  \bibfield  {author} {\bibinfo {author} {\bibfnamefont {W.}~\bibnamefont {K{\"u}rner}}, \bibinfo {author} {\bibfnamefont {R.}~\bibnamefont {Sauer}}, \bibinfo {author} {\bibfnamefont {A.}~\bibnamefont {D{\"o}rnen}},\ and\ \bibinfo {author} {\bibfnamefont {K.}~\bibnamefont {Thonke}},\ }\bibfield  {title} {\bibinfo {title} {Structure of the 0.767-ev oxygen-carbon luminescence defect in 450 c thermally annealed czochralski-grown silicon},\ }\href@noop {} {\bibfield  {journal} {\bibinfo  {journal} {Physical Review B}\ }\textbf {\bibinfo {volume} {39}},\ \bibinfo {pages} {13327} (\bibinfo {year} {1989})}\BibitemShut {NoStop}%
\bibitem [{\citenamefont {Davies}\ \emph {et~al.}(1989{\natexlab{a}})\citenamefont {Davies}, \citenamefont {Brian}, \citenamefont {Lightowlers}, \citenamefont {Barraclough},\ and\ \citenamefont {Thomaz}}]{davies1989temperature}%
  \BibitemOpen
  \bibfield  {author} {\bibinfo {author} {\bibfnamefont {G.}~\bibnamefont {Davies}}, \bibinfo {author} {\bibfnamefont {H.}~\bibnamefont {Brian}}, \bibinfo {author} {\bibfnamefont {E.}~\bibnamefont {Lightowlers}}, \bibinfo {author} {\bibfnamefont {K.}~\bibnamefont {Barraclough}},\ and\ \bibinfo {author} {\bibfnamefont {M.}~\bibnamefont {Thomaz}},\ }\bibfield  {title} {\bibinfo {title} {The temperature dependence of the 969 mev'g'optical transition in silicon},\ }\href@noop {} {\bibfield  {journal} {\bibinfo  {journal} {Semiconductor Science and Technology}\ }\textbf {\bibinfo {volume} {4}},\ \bibinfo {pages} {200} (\bibinfo {year} {1989}{\natexlab{a}})}\BibitemShut {NoStop}%
\bibitem [{\citenamefont {Davies}\ \emph {et~al.}(2006)\citenamefont {Davies}, \citenamefont {Hayama}, \citenamefont {Murin}, \citenamefont {Krause-Rehberg}, \citenamefont {Bondarenko}, \citenamefont {Sengupta}, \citenamefont {Davia},\ and\ \citenamefont {Karpenko}}]{Davies_Hayama_Murin_2006}%
  \BibitemOpen
  \bibfield  {author} {\bibinfo {author} {\bibfnamefont {G.}~\bibnamefont {Davies}}, \bibinfo {author} {\bibfnamefont {S.}~\bibnamefont {Hayama}}, \bibinfo {author} {\bibfnamefont {L.}~\bibnamefont {Murin}}, \bibinfo {author} {\bibfnamefont {R.}~\bibnamefont {Krause-Rehberg}}, \bibinfo {author} {\bibfnamefont {V.}~\bibnamefont {Bondarenko}}, \bibinfo {author} {\bibfnamefont {A.}~\bibnamefont {Sengupta}}, \bibinfo {author} {\bibfnamefont {C.}~\bibnamefont {Davia}},\ and\ \bibinfo {author} {\bibfnamefont {A.}~\bibnamefont {Karpenko}},\ }\bibfield  {title} {\bibinfo {title} {Radiation damage in silicon exposed to high-energy protons},\ }\href@noop {} {\bibfield  {journal} {\bibinfo  {journal} {Physical Review B—Condensed Matter and Materials Physics}\ }\textbf {\bibinfo {volume} {73}},\ \bibinfo {pages} {165202} (\bibinfo {year} {2006})}\BibitemShut {NoStop}%
\bibitem [{\citenamefont {Hall{\'e}n}\ \emph {et~al.}(1996)\citenamefont {Hall{\'e}n}, \citenamefont {Keskitalo}, \citenamefont {Masszi},\ and\ \citenamefont {N{\'a}gl}}]{hallen1996lifetime}%
  \BibitemOpen
  \bibfield  {author} {\bibinfo {author} {\bibfnamefont {A.}~\bibnamefont {Hall{\'e}n}}, \bibinfo {author} {\bibfnamefont {N.}~\bibnamefont {Keskitalo}}, \bibinfo {author} {\bibfnamefont {F.}~\bibnamefont {Masszi}},\ and\ \bibinfo {author} {\bibfnamefont {V.}~\bibnamefont {N{\'a}gl}},\ }\bibfield  {title} {\bibinfo {title} {Lifetime in proton irradiated silicon},\ }\href@noop {} {\bibfield  {journal} {\bibinfo  {journal} {Journal of Applied Physics}\ }\textbf {\bibinfo {volume} {79}},\ \bibinfo {pages} {3906} (\bibinfo {year} {1996})}\BibitemShut {NoStop}%
\bibitem [{\citenamefont {Biersack}\ and\ \citenamefont {Haggmark}(1980)}]{biersack80}%
  \BibitemOpen
  \bibfield  {author} {\bibinfo {author} {\bibfnamefont {J.~P.}\ \bibnamefont {Biersack}}\ and\ \bibinfo {author} {\bibfnamefont {L.~G.}\ \bibnamefont {Haggmark}},\ }\bibfield  {title} {\bibinfo {title} {A {Monte} {Carlo} computer program for the transport of energetic ions in amorphous targets},\ }\href {https://doi.org/10.1016/0029-554X(80)90440-1} {\bibfield  {journal} {\bibinfo  {journal} {Nuclear Instruments and Methods}\ }\textbf {\bibinfo {volume} {174}},\ \bibinfo {pages} {257} (\bibinfo {year} {1980})}\BibitemShut {NoStop}%
\bibitem [{\citenamefont {Ziegler}(1999)}]{ziegler1999stopping}%
  \BibitemOpen
  \bibfield  {author} {\bibinfo {author} {\bibfnamefont {J.~F.}\ \bibnamefont {Ziegler}},\ }\bibfield  {title} {\bibinfo {title} {Stopping of energetic light ions in elemental matter},\ }\href@noop {} {\bibfield  {journal} {\bibinfo  {journal} {Journal of applied physics}\ }\textbf {\bibinfo {volume} {85}},\ \bibinfo {pages} {1249} (\bibinfo {year} {1999})}\BibitemShut {NoStop}%
\bibitem [{\citenamefont {Agarwal}\ \emph {et~al.}(2021)\citenamefont {Agarwal}, \citenamefont {Lin}, \citenamefont {Li}, \citenamefont {Stoller},\ and\ \citenamefont {Zinkle}}]{Agarwal_Lin_Li_Stoller_Zinkle_2021}%
  \BibitemOpen
  \bibfield  {author} {\bibinfo {author} {\bibfnamefont {S.}~\bibnamefont {Agarwal}}, \bibinfo {author} {\bibfnamefont {Y.}~\bibnamefont {Lin}}, \bibinfo {author} {\bibfnamefont {C.}~\bibnamefont {Li}}, \bibinfo {author} {\bibfnamefont {R.~E.}\ \bibnamefont {Stoller}},\ and\ \bibinfo {author} {\bibfnamefont {S.~J.}\ \bibnamefont {Zinkle}},\ }\bibfield  {title} {\bibinfo {title} {On the use of srim for calculating vacancy production: Quick calculation and full-cascades options},\ }\href@noop {} {\bibfield  {journal} {\bibinfo  {journal} {Nuclear Instruments and Methods in Physics Research B}\ }\textbf {\bibinfo {volume} {503}},\ \bibinfo {pages} {11–29} (\bibinfo {year} {2021})}\BibitemShut {NoStop}%
\bibitem [{\citenamefont {Achilli}\ \emph {et~al.}(2025)\citenamefont {Achilli}, \citenamefont {Marian}, \citenamefont {Lodari}, \citenamefont {Bonera}, \citenamefont {Scappucci}, \citenamefont {Pedrini}, \citenamefont {Virgilio},\ and\ \citenamefont {Pezzoli}}]{achilli_25}%
  \BibitemOpen
  \bibfield  {author} {\bibinfo {author} {\bibfnamefont {S.}~\bibnamefont {Achilli}}, \bibinfo {author} {\bibfnamefont {D.}~\bibnamefont {Marian}}, \bibinfo {author} {\bibfnamefont {M.}~\bibnamefont {Lodari}}, \bibinfo {author} {\bibfnamefont {E.}~\bibnamefont {Bonera}}, \bibinfo {author} {\bibfnamefont {G.}~\bibnamefont {Scappucci}}, \bibinfo {author} {\bibfnamefont {J.}~\bibnamefont {Pedrini}}, \bibinfo {author} {\bibfnamefont {M.}~\bibnamefont {Virgilio}},\ and\ \bibinfo {author} {\bibfnamefont {F.}~\bibnamefont {Pezzoli}},\ }\href {https://doi.org/10.48550/arXiv.2506.09926} {\bibinfo {title} {Optical spin pumping in silicon}} (\bibinfo {year} {2025})\BibitemShut {NoStop}%
\bibitem [{\citenamefont {Tajima}(1977)}]{tajima1977photoluminescence}%
  \BibitemOpen
  \bibfield  {author} {\bibinfo {author} {\bibfnamefont {M.}~\bibnamefont {Tajima}},\ }\bibfield  {title} {\bibinfo {title} {Photoluminescence analyses of shallow impurities in silicon},\ }\href@noop {} {\bibfield  {journal} {\bibinfo  {journal} {Japanese Journal of Applied Physics}\ }\textbf {\bibinfo {volume} {16}},\ \bibinfo {pages} {2263} (\bibinfo {year} {1977})}\BibitemShut {NoStop}%
\bibitem [{\citenamefont {Karaiskaj}\ \emph {et~al.}(2001)\citenamefont {Karaiskaj}, \citenamefont {Thewalt}, \citenamefont {Ruf}, \citenamefont {Cardona}, \citenamefont {Pohl}, \citenamefont {Deviatych}, \citenamefont {Sennikov},\ and\ \citenamefont {Riemann}}]{PhysRevLett.86.6010}%
  \BibitemOpen
  \bibfield  {author} {\bibinfo {author} {\bibfnamefont {D.}~\bibnamefont {Karaiskaj}}, \bibinfo {author} {\bibfnamefont {M.~L.~W.}\ \bibnamefont {Thewalt}}, \bibinfo {author} {\bibfnamefont {T.}~\bibnamefont {Ruf}}, \bibinfo {author} {\bibfnamefont {M.}~\bibnamefont {Cardona}}, \bibinfo {author} {\bibfnamefont {H.-J.}\ \bibnamefont {Pohl}}, \bibinfo {author} {\bibfnamefont {G.~G.}\ \bibnamefont {Deviatych}}, \bibinfo {author} {\bibfnamefont {P.~G.}\ \bibnamefont {Sennikov}},\ and\ \bibinfo {author} {\bibfnamefont {H.}~\bibnamefont {Riemann}},\ }\bibfield  {title} {\bibinfo {title} {Photoluminescence of isotopically purified silicon: How sharp are bound exciton transitions?},\ }\href {https://doi.org/10.1103/PhysRevLett.86.6010} {\bibfield  {journal} {\bibinfo  {journal} {Phys. Rev. Lett.}\ }\textbf {\bibinfo {volume} {86}},\ \bibinfo {pages} {6010} (\bibinfo {year} {2001})}\BibitemShut {NoStop}%
\bibitem [{\citenamefont {Wagner}\ \emph {et~al.}(1984)\citenamefont {Wagner}, \citenamefont {Thonke},\ and\ \citenamefont {Sauer}}]{Wagner_Thonke_Sauer_1984}%
  \BibitemOpen
  \bibfield  {author} {\bibinfo {author} {\bibfnamefont {J.}~\bibnamefont {Wagner}}, \bibinfo {author} {\bibfnamefont {K.}~\bibnamefont {Thonke}},\ and\ \bibinfo {author} {\bibfnamefont {R.}~\bibnamefont {Sauer}},\ }\bibfield  {title} {\bibinfo {title} {Excitation spectroscopy on the 0.79-ev (c) line defect in irradiated silicon},\ }\href@noop {} {\bibfield  {journal} {\bibinfo  {journal} {Physical Review B}\ }\textbf {\bibinfo {volume} {29}},\ \bibinfo {pages} {7051} (\bibinfo {year} {1984})}\BibitemShut {NoStop}%
\bibitem [{\citenamefont {Thonke}\ \emph {et~al.}(1985)\citenamefont {Thonke}, \citenamefont {Wagner}, \citenamefont {Hangleiter},\ and\ \citenamefont {Sauer}}]{Thonke_Wagner_Hangleiter_Sauer_1985}%
  \BibitemOpen
  \bibfield  {author} {\bibinfo {author} {\bibfnamefont {K.}~\bibnamefont {Thonke}}, \bibinfo {author} {\bibfnamefont {J.}~\bibnamefont {Wagner}}, \bibinfo {author} {\bibfnamefont {A.}~\bibnamefont {Hangleiter}},\ and\ \bibinfo {author} {\bibfnamefont {R.}~\bibnamefont {Sauer}},\ }\bibfield  {title} {\bibinfo {title} {0.79 ev (c line) defect in irradiated oxygen-rich silicon: excited state structure, internal strain and luminescence decay time},\ }\href@noop {} {\bibfield  {journal} {\bibinfo  {journal} {Journal of Physics C: Solid State Physics}\ }\textbf {\bibinfo {volume} {18}},\ \bibinfo {pages} {L795–L801} (\bibinfo {year} {1985})}\BibitemShut {NoStop}%
\bibitem [{\citenamefont {Silkinis}\ \emph {et~al.}(2025)\citenamefont {Silkinis}, \citenamefont {Maciaszek}, \citenamefont {\ifmmode~\check{Z}\else \v{Z}\fi{}alandauskas}, \citenamefont {Bathen}, \citenamefont {Vines}, \citenamefont {Alkauskas},\ and\ \citenamefont {Razinkovas}}]{Silkinis25}%
  \BibitemOpen
  \bibfield  {author} {\bibinfo {author} {\bibfnamefont {R.}~\bibnamefont {Silkinis}}, \bibinfo {author} {\bibfnamefont {M.}~\bibnamefont {Maciaszek}}, \bibinfo {author} {\bibfnamefont {V.}~\bibnamefont {\ifmmode~\check{Z}\else \v{Z}\fi{}alandauskas}}, \bibinfo {author} {\bibfnamefont {M.~E.}\ \bibnamefont {Bathen}}, \bibinfo {author} {\bibfnamefont {L.}~\bibnamefont {Vines}}, \bibinfo {author} {\bibfnamefont {A.}~\bibnamefont {Alkauskas}},\ and\ \bibinfo {author} {\bibfnamefont {L.}~\bibnamefont {Razinkovas}},\ }\bibfield  {title} {\bibinfo {title} {Optical lineshapes of the c center in silicon from ab initio calculations: Interplay of localized modes and bulk phonons},\ }\href {https://doi.org/10.1103/PhysRevB.111.125136} {\bibfield  {journal} {\bibinfo  {journal} {Phys. Rev. B}\ }\textbf {\bibinfo {volume} {111}},\ \bibinfo {pages} {125136} (\bibinfo {year} {2025})}\BibitemShut {NoStop}%
\bibitem [{\citenamefont {Udvarhelyi}\ \emph {et~al.}(2022)\citenamefont {Udvarhelyi}, \citenamefont {Pershin}, \citenamefont {De{\'a}k},\ and\ \citenamefont {Gali}}]{udvarhelyi2022band}%
  \BibitemOpen
  \bibfield  {author} {\bibinfo {author} {\bibfnamefont {P.}~\bibnamefont {Udvarhelyi}}, \bibinfo {author} {\bibfnamefont {A.}~\bibnamefont {Pershin}}, \bibinfo {author} {\bibfnamefont {P.}~\bibnamefont {De{\'a}k}},\ and\ \bibinfo {author} {\bibfnamefont {A.}~\bibnamefont {Gali}},\ }\bibfield  {title} {\bibinfo {title} {An l-band emitter with quantum memory in silicon},\ }\href@noop {} {\bibfield  {journal} {\bibinfo  {journal} {npj Computational Materials}\ }\textbf {\bibinfo {volume} {8}},\ \bibinfo {pages} {262} (\bibinfo {year} {2022})}\BibitemShut {NoStop}%
\bibitem [{\citenamefont {Liu}\ \emph {et~al.}(2015)\citenamefont {Liu}, \citenamefont {Gao},\ and\ \citenamefont {Kakimoto}}]{LIU201558}%
  \BibitemOpen
  \bibfield  {author} {\bibinfo {author} {\bibfnamefont {X.}~\bibnamefont {Liu}}, \bibinfo {author} {\bibfnamefont {B.}~\bibnamefont {Gao}},\ and\ \bibinfo {author} {\bibfnamefont {K.}~\bibnamefont {Kakimoto}},\ }\bibfield  {title} {\bibinfo {title} {Numerical investigation of carbon contamination during the melting process of czochralski silicon crystal growth},\ }\href {https://doi.org/https://doi.org/10.1016/j.jcrysgro.2014.07.040} {\bibfield  {journal} {\bibinfo  {journal} {Journal of Crystal Growth}\ }\textbf {\bibinfo {volume} {417}},\ \bibinfo {pages} {58} (\bibinfo {year} {2015})}\BibitemShut {NoStop}%
\bibitem [{\citenamefont {Lappa}(2004)}]{LAPPA20041}%
  \BibitemOpen
  \bibfield  {author} {\bibinfo {author} {\bibfnamefont {M.}~\bibnamefont {Lappa}},\ }\bibfield  {title} {\bibinfo {title} {Chapter 1 - space research},\ }in\ \href {https://doi.org/https://doi.org/10.1016/B978-008044508-3/50002-5} {\emph {\bibinfo {booktitle} {Fluids, Materials and Microgravity}}}\ (\bibinfo  {publisher} {Elsevier},\ \bibinfo {address} {Oxford},\ \bibinfo {year} {2004})\ pp.\ \bibinfo {pages} {1--37}\BibitemShut {NoStop}%
\bibitem [{\citenamefont {Newman}(2000)}]{newman2000oxygen}%
  \BibitemOpen
  \bibfield  {author} {\bibinfo {author} {\bibfnamefont {R.}~\bibnamefont {Newman}},\ }\bibfield  {title} {\bibinfo {title} {Oxygen diffusion and precipitation in czochralski silicon},\ }\href@noop {} {\bibfield  {journal} {\bibinfo  {journal} {Journal of Physics: Condensed Matter}\ }\textbf {\bibinfo {volume} {12}},\ \bibinfo {pages} {R335} (\bibinfo {year} {2000})}\BibitemShut {NoStop}%
\bibitem [{\citenamefont {Nagai}\ \emph {et~al.}(2014)\citenamefont {Nagai}, \citenamefont {Nakagawa},\ and\ \citenamefont {Kashima}}]{NAGAI2014737}%
  \BibitemOpen
  \bibfield  {author} {\bibinfo {author} {\bibfnamefont {Y.}~\bibnamefont {Nagai}}, \bibinfo {author} {\bibfnamefont {S.}~\bibnamefont {Nakagawa}},\ and\ \bibinfo {author} {\bibfnamefont {K.}~\bibnamefont {Kashima}},\ }\bibfield  {title} {\bibinfo {title} {Crystal growth of mcz silicon with ultralow carbon concentration},\ }\href {https://doi.org/https://doi.org/10.1016/j.jcrysgro.2013.11.059} {\bibfield  {journal} {\bibinfo  {journal} {Journal of Crystal Growth}\ }\textbf {\bibinfo {volume} {401}},\ \bibinfo {pages} {737} (\bibinfo {year} {2014})},\ \bibinfo {note} {proceedings of 17th International Conference on Crystal Growth and Epitaxy (ICCGE-17)}\BibitemShut {NoStop}%
\bibitem [{\citenamefont {Müller}\ and\ \citenamefont {Friedrich}(2005)}]{MULLER2005262}%
  \BibitemOpen
  \bibfield  {author} {\bibinfo {author} {\bibfnamefont {G.}~\bibnamefont {Müller}}\ and\ \bibinfo {author} {\bibfnamefont {J.}~\bibnamefont {Friedrich}},\ }\bibfield  {title} {\bibinfo {title} {Crystal growth, bulk: Methods},\ }in\ \href {https://doi.org/https://doi.org/10.1016/B0-12-369401-9/00416-2} {\emph {\bibinfo {booktitle} {Encyclopedia of Condensed Matter Physics}}}\ (\bibinfo  {publisher} {Elsevier},\ \bibinfo {address} {Oxford},\ \bibinfo {year} {2005})\ pp.\ \bibinfo {pages} {262--274}\BibitemShut {NoStop}%
\bibitem [{\citenamefont {Schutz-Kuchly}\ \emph {et~al.}(2010)\citenamefont {Schutz-Kuchly}, \citenamefont {Veirman}, \citenamefont {Dubois},\ and\ \citenamefont {Heslinga}}]{schutz2010light}%
  \BibitemOpen
  \bibfield  {author} {\bibinfo {author} {\bibfnamefont {T.}~\bibnamefont {Schutz-Kuchly}}, \bibinfo {author} {\bibfnamefont {J.}~\bibnamefont {Veirman}}, \bibinfo {author} {\bibfnamefont {S.}~\bibnamefont {Dubois}},\ and\ \bibinfo {author} {\bibfnamefont {D.}~\bibnamefont {Heslinga}},\ }\bibfield  {title} {\bibinfo {title} {Light-induced-degradation effects in boron--phosphorus compensated n-type czochralski silicon},\ }\href@noop {} {\bibfield  {journal} {\bibinfo  {journal} {Applied Physics Letters}\ }\textbf {\bibinfo {volume} {96}} (\bibinfo {year} {2010})}\BibitemShut {NoStop}%
\bibitem [{\citenamefont {Schmidt}\ and\ \citenamefont {Bothe}(2004)}]{schmidt2004structure}%
  \BibitemOpen
  \bibfield  {author} {\bibinfo {author} {\bibfnamefont {J.}~\bibnamefont {Schmidt}}\ and\ \bibinfo {author} {\bibfnamefont {K.}~\bibnamefont {Bothe}},\ }\bibfield  {title} {\bibinfo {title} {Structure and transformation of the metastable boron-and oxygen-related defect center in crystalline silicon},\ }\href@noop {} {\bibfield  {journal} {\bibinfo  {journal} {Physical review B}\ }\textbf {\bibinfo {volume} {69}},\ \bibinfo {pages} {024107} (\bibinfo {year} {2004})}\BibitemShut {NoStop}%
\bibitem [{\citenamefont {P{\"a}ssler}(1997)}]{passler1997basic}%
  \BibitemOpen
  \bibfield  {author} {\bibinfo {author} {\bibfnamefont {R.}~\bibnamefont {P{\"a}ssler}},\ }\bibfield  {title} {\bibinfo {title} {Basic model relations for temperature dependencies of fundamental energy gaps in semiconductors},\ }\href@noop {} {\bibfield  {journal} {\bibinfo  {journal} {physica status solidi (b)}\ }\textbf {\bibinfo {volume} {200}},\ \bibinfo {pages} {155} (\bibinfo {year} {1997})}\BibitemShut {NoStop}%
\bibitem [{\citenamefont {Davies}\ \emph {et~al.}(1989{\natexlab{b}})\citenamefont {Davies}, \citenamefont {Brian}, \citenamefont {Lightowlers}, \citenamefont {Barraclough},\ and\ \citenamefont {Thomaz}}]{Davies_Brian_Lightowlers_Barraclough_THomaz_1989}%
  \BibitemOpen
  \bibfield  {author} {\bibinfo {author} {\bibfnamefont {G.}~\bibnamefont {Davies}}, \bibinfo {author} {\bibfnamefont {H.}~\bibnamefont {Brian}}, \bibinfo {author} {\bibfnamefont {E.~C.}\ \bibnamefont {Lightowlers}}, \bibinfo {author} {\bibfnamefont {K.}~\bibnamefont {Barraclough}},\ and\ \bibinfo {author} {\bibfnamefont {M.~F.}\ \bibnamefont {Thomaz}},\ }\bibfield  {title} {\bibinfo {title} {The temperature dependence of the 969 mev “g” optical transition in silicon},\ }\href@noop {} {\bibfield  {journal} {\bibinfo  {journal} {Semiconductor Science and Technology}\ }\textbf {\bibinfo {volume} {4}},\ \bibinfo {pages} {200–206} (\bibinfo {year} {1989}{\natexlab{b}})}\BibitemShut {NoStop}%
\bibitem [{\citenamefont {Streetman}\ \emph {et~al.}(1971)\citenamefont {Streetman}, \citenamefont {Compton}, \citenamefont {Johnson}, \citenamefont {Jones}, \citenamefont {Noonan},\ and\ \citenamefont {Wong}}]{Streetman_Compton_Johnson_Jones_Noonan_Wong_1971}%
  \BibitemOpen
  \bibfield  {author} {\bibinfo {author} {\bibfnamefont {B.~G.}\ \bibnamefont {Streetman}}, \bibinfo {author} {\bibfnamefont {W.~D.}\ \bibnamefont {Compton}}, \bibinfo {author} {\bibfnamefont {E.~S.}\ \bibnamefont {Johnson}}, \bibinfo {author} {\bibfnamefont {C.~E.}\ \bibnamefont {Jones}}, \bibinfo {author} {\bibfnamefont {J.~R.}\ \bibnamefont {Noonan}},\ and\ \bibinfo {author} {\bibfnamefont {E.~S.}\ \bibnamefont {Wong}},\ }\href@noop {} {\emph {\bibinfo {title} {A study of irradiation-induced defects in Silicon using low temperature photoluminescence}}}\ (\bibinfo {year} {1971})\BibitemShut {NoStop}%
\bibitem [{\citenamefont {Jones}\ \emph {et~al.}(1973{\natexlab{b}})\citenamefont {Jones}, \citenamefont {Johnson}, \citenamefont {Compton}, \citenamefont {Noonan},\ and\ \citenamefont {Streetman}}]{Jones_Johnson_Compton_Noonan_Streetman_1973}%
  \BibitemOpen
  \bibfield  {author} {\bibinfo {author} {\bibfnamefont {C.~E.}\ \bibnamefont {Jones}}, \bibinfo {author} {\bibfnamefont {E.~S.}\ \bibnamefont {Johnson}}, \bibinfo {author} {\bibfnamefont {W.~D.}\ \bibnamefont {Compton}}, \bibinfo {author} {\bibfnamefont {J.~R.}\ \bibnamefont {Noonan}},\ and\ \bibinfo {author} {\bibfnamefont {B.~G.}\ \bibnamefont {Streetman}},\ }\bibfield  {title} {\bibinfo {title} {Temperature, stress, and annealing effects on the luminescence from electron-irradiated silicon},\ }\href@noop {} {\bibfield  {journal} {\bibinfo  {journal} {Journal of Applied Physics}\ }\textbf {\bibinfo {volume} {44}},\ \bibinfo {pages} {5402–5410} (\bibinfo {year} {1973}{\natexlab{b}})}\BibitemShut {NoStop}%
\bibitem [{\citenamefont {Quard}\ \emph {et~al.}(2024)\citenamefont {Quard}, \citenamefont {Khoury}, \citenamefont {Wang}, \citenamefont {Herzig}, \citenamefont {Meijer}, \citenamefont {Pezzagna}, \citenamefont {Cueff}, \citenamefont {Grojo}, \citenamefont {Abbarchi}, \citenamefont {Nguyen} \emph {et~al.}}]{quard2024femtosecond}%
  \BibitemOpen
  \bibfield  {author} {\bibinfo {author} {\bibfnamefont {H.}~\bibnamefont {Quard}}, \bibinfo {author} {\bibfnamefont {M.}~\bibnamefont {Khoury}}, \bibinfo {author} {\bibfnamefont {A.}~\bibnamefont {Wang}}, \bibinfo {author} {\bibfnamefont {T.}~\bibnamefont {Herzig}}, \bibinfo {author} {\bibfnamefont {J.}~\bibnamefont {Meijer}}, \bibinfo {author} {\bibfnamefont {S.}~\bibnamefont {Pezzagna}}, \bibinfo {author} {\bibfnamefont {S.}~\bibnamefont {Cueff}}, \bibinfo {author} {\bibfnamefont {D.}~\bibnamefont {Grojo}}, \bibinfo {author} {\bibfnamefont {M.}~\bibnamefont {Abbarchi}}, \bibinfo {author} {\bibfnamefont {H.~S.}\ \bibnamefont {Nguyen}}, \emph {et~al.},\ }\bibfield  {title} {\bibinfo {title} {Femtosecond-laser-induced creation of g and w color centers in silicon-on-insulator substrates},\ }\href@noop {} {\bibfield  {journal} {\bibinfo  {journal} {Physical Review Applied}\ }\textbf {\bibinfo {volume} {21}},\ \bibinfo {pages} {044014} (\bibinfo {year} {2024})}\BibitemShut {NoStop}%
\bibitem [{\citenamefont {Vines}\ \emph {et~al.}(2019)\citenamefont {Vines}, \citenamefont {Kuzmenko}, \citenamefont {Kirdoda}, \citenamefont {Dumas}, \citenamefont {Mirza}, \citenamefont {Millar}, \citenamefont {Paul},\ and\ \citenamefont {Buller}}]{vines_2019}%
  \BibitemOpen
  \bibfield  {author} {\bibinfo {author} {\bibfnamefont {P.}~\bibnamefont {Vines}}, \bibinfo {author} {\bibfnamefont {K.}~\bibnamefont {Kuzmenko}}, \bibinfo {author} {\bibfnamefont {J.}~\bibnamefont {Kirdoda}}, \bibinfo {author} {\bibfnamefont {D.~C.~S.}\ \bibnamefont {Dumas}}, \bibinfo {author} {\bibfnamefont {M.~M.}\ \bibnamefont {Mirza}}, \bibinfo {author} {\bibfnamefont {R.~W.}\ \bibnamefont {Millar}}, \bibinfo {author} {\bibfnamefont {D.~J.}\ \bibnamefont {Paul}},\ and\ \bibinfo {author} {\bibfnamefont {G.~S.}\ \bibnamefont {Buller}},\ }\bibfield  {title} {{\bibinfo {title} {High performance planar germanium-on-silicon single-photon avalanche diode detectors}},\ }\href {https://doi.org/10.1038/s41467-019-08830-w} {\bibfield  {journal} {\bibinfo  {journal} {Nature Communications}\ }\textbf {\bibinfo {volume} {10}},\ \bibinfo {pages} {1086} (\bibinfo {year} {2019})}\BibitemShut {NoStop}%
\bibitem [{\citenamefont {Na}\ \emph {et~al.}(2024)\citenamefont {Na}, \citenamefont {Lu}, \citenamefont {Liu}, \citenamefont {Chen}, \citenamefont {Lai}, \citenamefont {Lin}, \citenamefont {Lin}, \citenamefont {Shia}, \citenamefont {Cheng},\ and\ \citenamefont {Chen}}]{na_2024}%
  \BibitemOpen
  \bibfield  {author} {\bibinfo {author} {\bibfnamefont {N.}~\bibnamefont {Na}}, \bibinfo {author} {\bibfnamefont {Y.-C.}\ \bibnamefont {Lu}}, \bibinfo {author} {\bibfnamefont {Y.-H.}\ \bibnamefont {Liu}}, \bibinfo {author} {\bibfnamefont {P.-W.}\ \bibnamefont {Chen}}, \bibinfo {author} {\bibfnamefont {Y.-C.}\ \bibnamefont {Lai}}, \bibinfo {author} {\bibfnamefont {Y.-R.}\ \bibnamefont {Lin}}, \bibinfo {author} {\bibfnamefont {C.-C.}\ \bibnamefont {Lin}}, \bibinfo {author} {\bibfnamefont {T.}~\bibnamefont {Shia}}, \bibinfo {author} {\bibfnamefont {C.-H.}\ \bibnamefont {Cheng}},\ and\ \bibinfo {author} {\bibfnamefont {S.-L.}\ \bibnamefont {Chen}},\ }\bibfield  {title} {{\bibinfo {title} {Room temperature operation of germanium–silicon single-photon avalanche diode}},\ }\href {https://doi.org/10.1038/s41586-024-07076-x} {\bibfield  {journal} {\bibinfo  {journal} {Nature}\ }\textbf {\bibinfo {volume} {627}},\ \bibinfo {pages} {295} (\bibinfo {year} {2024})}\BibitemShut {NoStop}%
\bibitem [{\citenamefont {Huygens}\ \emph {et~al.}(2007)\citenamefont {Huygens}, \citenamefont {Gomes},\ and\ \citenamefont {Strubbe}}]{huygens2007etching}%
  \BibitemOpen
  \bibfield  {author} {\bibinfo {author} {\bibfnamefont {I.~M.}\ \bibnamefont {Huygens}}, \bibinfo {author} {\bibfnamefont {W.}~\bibnamefont {Gomes}},\ and\ \bibinfo {author} {\bibfnamefont {K.}~\bibnamefont {Strubbe}},\ }\bibfield  {title} {\bibinfo {title} {Etching of germanium in hydrogenperoxide solutions},\ }\href@noop {} {\bibfield  {journal} {\bibinfo  {journal} {ECS Transactions}\ }\textbf {\bibinfo {volume} {6}},\ \bibinfo {pages} {375} (\bibinfo {year} {2007})}\BibitemShut {NoStop}%
\end{thebibliography}
\end{document}